\documentclass[pra,superscriptadress]{revtex4}
\usepackage{graphicx}
\usepackage{amsmath}
\usepackage{amssymb}
\usepackage{bm}

\begin{document}

\title{Optimal conclusive teleportation of quantum states}
\author{L. Roa$^{1,2}$, A. Delgado$^{1,2}$, and I. Fuentes-Guridi$^{3,4} $}
\affiliation{$^{1}$Department of Physics and Astronomy, University of New Mexico,
                   800 Yale Boulevard, 87131 Albuquerque, USA}
\affiliation{$^{2}$Departamento de F\' isica, Universidad de
Concepci\' on, casilla 160-C, Concepci\' on, Chile}
\affiliation{$^{3}$Perimeter Institute, 35 King Street, North
Waterloo, Ontario, Canada N2J 2W9} \affiliation{$^{4}$ Optics
Section, The Blackett Laboratory, Imperial College, London SW7
2BZ, United Kingdom}

\begin{abstract}
Quantum teleportation of qudits is revisited. In particular, we
analyze the case where the quantum channel corresponds to a
non-maximally entangled state and show that the success of the
protocol is directly related to the problem of distinguishing
non-orthogonal quantum states. The teleportation channel can be
seen as a coherent superposition of two channels, one of them
being a maximally entangled state thus, leading to perfect
teleportation and the other, corresponding to a non-maximally
entangled state living in a subspace of the $d$-dimensional
Hilbert space. The second channel leads to a teleported state with
reduced fidelity. We calculate the average fidelity of the process
and show its optimality.
\end{abstract}

\date{\today}
\maketitle

\section{introduction}

Entanglement is a fundamental property of quantum mechanical systems \cite{one}.
It is one of the most interesting and puzzling ideas associated with composite systems \cite{two}.
The postulates of quantum mechanics state that the state space of a composite physical system made up of two (or more) distinct
physical systems, is a tensor product of the state spaces of the component systems.
A consequence of this structure is that there are states in the composite state space for which the correlations
between the component systems cannot be accounted for classically.
By classical we mean here that only local operations and classical communications (LOCC) are considered.

Although there is still not a complete theory on entanglement, it
is considered a fundamental resource of nature whose importance is
comparable to energy, information and entropy \cite{three}, among
others. Recently, the field of entanglement has become an intense
research area due to its key role in many applications of quantum
information processing \cite{nielsen}. An important example of
this is teleportation of quantum states, where a maximally
entangled state shared by two parties is used as a channel to
transmit a unknown state using LOCCs. Teleportation protocols can
also be used to transmit quantum operations \cite{huelga1,huelga2}
and to implement protocols for quantum cryptography
\cite{Saavedra}.

In this paper we study the teleportation of a quantum state
belonging to a $d$-dimensional Hilbert space (qudit). Our protocol
considers the use of a non-maximally entangled pure state of two
qudits as quantum channel. We relate quantum teleportation to the
problem of quantum states discrimination and show that the success
of this scheme has a fundamental limit determined by how
accurately a set of non-orthogonal linearly independent quantum
states can be unambiguously distinguished. Thereby, we generalize
to arbitrary dimensions the result by T. Mor and P. Horodecki
\cite{Mor} concerning conclusive quantum teleportation of qubits.
We show that the generalized protocol for conclusive quantum
teleportation can be interpreted in terms of the coherent
superposition of two quantum channels. One of them allows for
perfect teleportation. The other channel corresponds to a
superposition of $d-1$ product states and leads to the failure of
the process. This interpretation of conclusive quantum
teleportation allow us to calculate easily the average fidelity
over the entire Hilbert space. Finally, we demonstrate the
optimality of the average fidelity.

This paper is organized as follows: In section \ref{sec:one} we
review the standard teleportation protocol considering both
maximally and non-maximally entangled states as quantum channel.
In this section we also relate quantum teleportation to quantum
state discrimination. In section \ref{sec:two} we discuss in
detail the quantum state discrimination protocol used in this
paper. The results of this section allows us to calculate in
section \ref{sec:three} the average fidelity of conclusive state
teleportation and demonstrate its optimality.

\section{Quantum state teleportation}
\label{sec:one}

In the process of teleporting  a quantum state two parties, sender
and receiver, share a maximally entangled two-qudit pure state.
The sender has a third qudit in the state to be teleported. The
sender carries out a generalized Bell measurement on his two
particles and communicates the outcome of the measurement to the
receiver. Conditional on the measurement
 result the receiver
applies an unitary transformation on his particle. Thereafter,
receiver's particle is in the state to be teleported.

The teleportation of a quantum state $|\psi\rangle$ of a
$d$-dimensional Hilbert space, spanned by the basis
$\{|n\rangle\}$ with $n=0,\dots, d-1$, can be shortly described by
the following identity
\begin{equation}
|\Psi_{0,0}\rangle_{12}\otimes|\psi\rangle_3=\frac{1}{d}
\sum_{l,k=0}^{d-1}Z_1^{d-l}X_1^k|\psi\rangle_1\otimes|\Psi_{l,k}\rangle_{23},
\label{ID1}
\end{equation}
where particles 2 and 3 belongs to the sender and particle 1 to
the receiver. The states $|\Psi_{l,k}\rangle$ with $l,k=0,\dots,
d-1$ are a generalization of the Bell basis to the case of two
$d$-dimensional quantum systems,
\begin{equation}
|\Psi_{n,m}\rangle=\frac{1}{\sqrt{d}} \sum_{j}e^{2\pi ijn/d
}|j\rangle\otimes|j\oplus m\rangle,
\end{equation}
where $j\oplus m$ denotes the sum $j+m$ modulus d.
 In this case the maximally
entangled state $|\Psi_{0,0}\rangle$ has been chosen as quantum
channel. The unitary operators $X$ and $Z$ are defined by
\begin{equation}
X=\sum_{n=0}^{d-1}|n+1\rangle\langle n|,~~~~~~Z=
\sum_{n=0}^{d-1}\exp(\frac{2\pi i}{d}n)|n\rangle\langle n|.
\end{equation}
Instead of a direct Bell measurement it is possible to apply a
generalized control-not gate $GXOR_{23}$ \cite{Alber} in order to
map the states $|\Psi_{l,k}\rangle_{23}$ onto the unentangled
states $F_2|l\rangle_2\otimes|k\rangle_3$, where $F$ denotes the
discrete Fourier transform, i.e.
\begin{equation}
F|l\rangle=\frac{1}{\sqrt{d}}\sum_{k=0}^{d-1}e^{i2\pi
lk/d}|k\rangle.
\end{equation}
The generalized control-not gate is defined by
$GXOR_{ab}|i\rangle_a|j\rangle_b=|i\rangle_a|i\ominus j\rangle$,
where $i\ominus j$ stands for the difference $i-j$ modulo d.
Thereby, Eq. (\ref{ID1}) becomes
\begin{equation}
GXOR_{23}|\Psi_{0,0}\rangle_{12}\otimes|\psi\rangle_3=\frac{1}{d}
\sum_{l,k=0}^{d-1}Z_1^{d-l}X_1^k|\psi\rangle_1\otimes
F_2|l\rangle_{2}\otimes|k\rangle_{3} .
\label{ID2}
\end{equation}
The protocol for quantum state teleportation can be straightforwardly read out from this equation.
In the case of a non-maximally entangled pure quantum state $|\Psi\rangle_{12}$ as quantum channel, defined by
\begin{equation}
|\Psi\rangle_{12}=\sum_{m=0}^{d-1}A_m|m\rangle_1\otimes|m\rangle_2
\label{Initial state}
\end{equation}
where $\{|m\rangle_i\}$ with  $i=1,2$ are orthonormal basis
defined by the Schmidt decomposition and the coefficients $A_m$
are real and satisfy the normalization condition, the previous
identity Eq. (\ref{ID2}) is replaced by
\begin{equation}
GXOR_{23}|\Psi\rangle_{12}\otimes|\psi\rangle_3=\frac{1}{d}
\sum_{l,k=0}^{d-1}Z_1^{d-l}X_1^k|\psi\rangle_1\otimes
|\nu_l\rangle_{2}\otimes|k\rangle_{3}
\label{ID4}
\end{equation}
where the states $|\nu_l\rangle_{2}$ are given by
\begin{equation}
|\nu_l\rangle_{2}=Z^l\sum_{k=0}^{d-1}A_k|k\rangle_2~~~~~\mathrm{with}%
~l=0\dots d-1.
\label{Nu}
\end{equation}
The previous identity Eq. (\ref{ID4}) resembles Eq. (\ref{ID2})
where now the states $F|l\rangle_2$  have been replaced by the
states $|\nu_l\rangle$ of Eq. (\ref{Nu}). Let us now recall that
in the teleportation of states it is necessary to measure the
state of particle 2. Conditional on the outcome of this
measurement a unitary operator is applied on particle one. These
operators are in a one to one relation with the outcomes of the
measurements. Therefore, it is necessary to distinguish the
possible states of particle 2 perfectly. However, it is clear from
the definition of the states $|\nu_l\rangle_{2}$ Eq. ({\ref{Nu}})
that in general these states are non-orthogonal. In fact, the
inner product between any two of these states is given by
$\langle\nu_n|\nu_m\rangle=\sum_{k=0}^{d-1}\exp(\frac{2\pi
i}{d}k(n-m))|A_k|^2$. Only in the case of a maximally entangled
state as quantum channel, that is
$A_k=1/\sqrt{d}~\forall~k=0,\dots,d-1$, the overlap vanishes and
the states $|\nu_l\rangle_{2}$ are simply the states
$F_2|l\rangle_2$ which are mutually orthogonal. Thus, in the case
of a non-maximally entangled state $|\psi\rangle_{12}$ the success
of the teleportation protocol is limited by our capability to
distinguish among the set $\{|\nu_l\rangle_{2}\}$ of $d$
non-orthogonal quantum states.

This problem has been previously studied by T. Mor and P.
Horodecky \cite{Mor}  in the case of two-dimensional quantum
systems. They proposed the use of unambiguous state discrimination
in combination with the usual quantum teleportation protocol. In
this way, it is possible to distinguish perfectly among the two
states $|\nu_0\rangle_2$ and $|\nu_1\rangle_2$ with some
probability. For those events in which the discrimination it is
successful the teleported state has fidelity one. However, if the
discrimination fails the post measurement states might still allow
to teleport thought with reduced fidelity.

In the next section we review briefly the problem of quantum state
discrimination. We show the optimal conclusive state
discrimination protocol for the states $|\nu_l\rangle_{2}$ Eq.
({\ref{Nu}}) and obtain the post measurement states.

\section{Quantum state discrimination}
\label{sec:two}

The problem of quantum state discrimination has deserved
considerable attention. An overview of the main strategies has
been given by Chefles \cite{Chefles00}. In the later, generalized
measurements are used to construct an error-free strategy for
discriminating among a finite number of non-orthogonal states with
given a priori probabilities. The scheme can occasionally lead to
inconclusive results. This idea first proposed by Ivanovic
\cite{ivanovic} has been studied by Dieks \cite{Dieks} and Peres
\cite{peres} for two non-orthogonal states generated with equal a
priory probabilities. The result was later generalized by Jaeger
and Shimony \cite{Jaeger} for arbitrary a priory probabilities.
The qudit case was then considered by Chefles \cite{Chefles98} and
Peres and Terno \cite{peres2} where the former showed that results
for the qudit case simply generalize form those of qubits when the
states are linearly independent. The linearly dependent case can
be considered only when copies of the state are available
\cite{Chefles01}. Here we consider a method developed by Sun et
al. \cite{Sun} which allows the construction of the optimal
conclusive state discrimination scheme for a given set of linearly
dependent states.

According to the quantum operations formalism \cite{Kraus} the
most  general transformation of a quantum system can be
represented by a completely positive, trace preserving map.
Thereby, it is possible to perform a predetermined non-unitary
transformation with some probability. Furthermore, in general it
is possible to known whether the desired transformation has been
successfully implemented or not.

In particular, it is possible to change probabilistically the
inner product between pure states of a quantum system. This is the
essence of unambiguous state discrimination among elements of a
set $\Omega=\{|\nu_l\rangle\}$ of non-orthogonal states.  The
state $|\nu_l\rangle$ is mapped probabilistically onto  the state
$|e_l\rangle$ which belongs to a set orthogonal states. This
mapping has certain probability of failure. In this case, the
system is mapped onto a state $|\phi_l\rangle$ which does not
allow a  conclusive identification of the initial state. Thus, the
failure probability of the mapping is identified with the total
probability of obtaining an inconclusive identification of the
states. A major problem consists in finding the optimal mapping,
that is the mapping with the smallest inconclusive probability.

Necessary and sufficient conditions for the existence of a
conclusive  discrimination scheme have been found by Chefles
\cite{Chefles}, namely the states in $\Omega$ must be linearly
independent (LI). Thus, the states $\{|\phi_l\rangle\}$ must be
linearly dependent. Otherwise it would be possible to use another
mapping which allows the discrimination among these states. Based
on this observation, Sun et al. \cite{Sun} have developed a method
to find the optimal conclusive discrimination scheme and proposed
a physical implementation in terms of optical multiports. In their
approach conclusive state discrimination is described in terms of
a unitary operator $U$ and projective measurements. The states
$\{|\nu_l\rangle\}$ generated with a priory probabilities
$\{\eta_l\}$ are considered to belong to a Hilbert space ${\cal
K}$ which can be decomposed as a  direct sum of two subspaces,
i.e. ${\cal K}={\cal U}\oplus{\cal A}$. These subspaces are
spanned  by the basis states $\{|u_l\rangle\}$ and
$\{|a_l\rangle\}$ respectively. The action of the unitary
transformation is such that
\begin{equation}
U|\nu_l\rangle=\sqrt{p_l}|u_l\rangle+|\phi_l\rangle,
\label{Unitarydiscrimination}
\end{equation}
where the set of not necessarily normalized, linearly dependent
states  $\{|\phi_l\rangle\}$ is in ${\cal A}$ and $p_l$ denotes
the probability of discriminating successfully the state
$|\nu_l\rangle$.  The unitary transformation $U$ Eq.
(\ref{Unitarydiscrimination}) is followed by a measurement which
projects the state of the particle onto one of the basis states of
$\{|u_l\rangle\}$ or $\{|a_l\rangle\}$.  The states
$|\nu_l\rangle$ and $|u_l\rangle$ are in one to one
correspondence. This and the orthogonality of the states
${|u_l\rangle}$ allows one to discriminate among the states of the
set $\{|\nu_l\rangle\}$. However, in general each basis state in
${\cal A}$ have a component in all the states $\{|\phi_l\rangle\}$
and thus it is not possible to assign them a particular state
$|\nu_l\rangle$.

The optimal average probability of succes
$S=\sum_{l=0}^{d-1}\eta_lp_l$ can be found under the constraint
$det(Q)=0$ where $Q$ is the positive semidefinite matrix whose
matrical elements are given by
\begin{equation}
Q_{k,l}=\langle\phi_k|\phi_l\rangle=\langle\nu_k|\nu_l\rangle-p_k\delta_{k,l}.
\end{equation}
The states $|\phi_l\rangle$ can be defined as $|\phi_l\rangle=A|a_l\rangle$ with $Q=A^\dagger A$.
Thereby, it is possible to find the form of $U$ in Eq. (\ref{Unitarydiscrimination}).

It turns out that the states $|\nu_l\rangle$ are not necessarily well suited for conclusive
discrimination. In fact, these states can be linearly dependent.  The states $|\nu_l\rangle$ are LI under
the condition
\begin{equation}
\sum_{l=0}^{d-1}C_l|\nu_l\rangle_{2}=0~~~~~\mathrm{iff}~C_l=0~\forall~l=0
\dots d-1
\end{equation}
or equivalently
\begin{equation}
\sum_{n=0}^{d-1}\langle c|F |n\rangle A_n|n\rangle=0~~~~~\mathrm{with}
~|c\rangle=\sum_{k=0}^{d-1}C_k^*|k\rangle.
\label{LILD}
\end{equation}
In the case that all the amplitudes $A_n$ are nonzero then all the coefficients
$\langle c|F |n\rangle$ must be null. Since $F |n\rangle$ form a basis the only
solution to Eq. (\ref{LILD}) is $|c\rangle=0$. Therefore, all the coefficients $C_l$
must vanish. Thus, if all the amplitudes $A_m$ are different from zero
the states $|\nu_l\rangle$ are LI. Otherwise, when a subset of the amplitudes $A_m$
are zero the Eq. (\ref{LILD}) can be satisfied by taking $|c\rangle=\sum_{\{m\}}a_mF|m\rangle$
for any $a_m\ne 0$. Thus, in this case the states $|\nu_l\rangle$ are LD.

In the case that all the amplitudes $A_m$ are different from zero all the states $|\nu_l\rangle$ are different.
Thus, the only source of error is the scheme of discrimination itself. For example, in the process of
conclusive state discrimination there is a probability for the failure in the discrimination process.
This event leads to a failure in the teleportation of states because it is not possible to decide
which unitary operator must be applied in order to recover the state to be teleported. When only
one of the amplitudes is different from zero the states $|\nu_l\rangle$ are all equal.
Henceforth, it is impossible to discriminate among them at all and both two processes fail completely.

Generally, when $1<n<d$ of the $d$ amplitudes $A_m$ are different from zero then the $d$
states $|\nu_l\rangle_{2}$ are LD. In this case it has been shown \cite{Chefles01} that
a conclusive state discrimination protocol can be formulated when copies for each state
$|\nu_l\rangle_{2}$ are available. This adds an extra source of error to the teleportation
of states and unitaries. In fact, the no cloning theorem \cite{Wootters} states that it
is not possible to copy perfectly an unknown quantum states due to the linear character of
quantum mechanics. Thus, besides the success probability of the discrimination protocol
itself the success probability of a probabilistic cloning machine must be considered.

Let us now calculate the optimal average failure probability $F=1-S$ for the states $\{|\nu_l\rangle\}$
under the condition $A_m\ne 0~\forall~m=0,\dots,d-1$. The calculations can be greatly
simplified if the matrix $Q$ is Fourier transformed. In fact, the matrix $Q$ becomes
\begin{equation}
F Q F^\dagger=\tilde Q= \sum_{m=0}^{d-1}\left [\left(\sum_{k=0}^{d-1}
\frac{f_k}{d}\right)-1+dA_m^2\right]|m\rangle\langle m| +\sum_{m\ne
n}^{d-1}\left(\sum_{k=0}^{d-1}\frac{f_k}{d}\epsilon^{k(n-m)}\right)|m\rangle
\langle n|,
\end{equation}
where $f_k$ denotes the failure probability asociated with the state $|\nu_k\rangle$.
For $d$ arbitrary the determinant of $\tilde Q$ has the generic form
\begin{equation}
\mathrm{det}(\tilde
Q)=C_0(A_0^2,\dots,A_{d-1}^2)+C_1(A_0^2,\dots,A_{d-1}^2)\sum_{k=0}^{d-1}f_k+\Pi_{k=0}^{d-1}f_k.
\end{equation}
Applying the method of Lagrange multipliers we obtain
\begin{equation}
\frac{\partial}{\partial f_i}\left(F+\lambda \mathrm{det}(\tilde Q)\right)=%
\frac{1}{d}+\lambda C_1(A_0^2,\dots,A_{d-1}^2)+\lambda\Pi_{k\ne
i}^{d-1}f_k.
\label{Derivatives}
\end{equation}
where $\lambda$ is the Lagrange multiplier and we have made use of the fact that the
states $\{|\nu_l\rangle\}$ are generated with the same probability, i.e. $\eta_l=1/d~\forall~l=0,\dots,d-1$.
The derivatives Eq. (\ref{Derivatives}) are invariant under permutations of the $ f_i $'s.
In particular, they can be obtained from the derivative with respect to $f_0$ by suitably
permuting the $f_i^{\prime}s$. Thereby, the condition
\begin{equation}
\frac{\partial}{\partial f_i}\left(F+\lambda \mathrm{det}(\tilde Q)\right)=0
\end{equation}
implies that $f_0=f_1=\dots=f_{d-1}=f$. Thus, the failure probabilities are all equal.
Under this condition the matrix $\tilde Q$ has the simpler expression
\begin{equation}
\tilde Q= \sum_{m=0}^{d-1}\left(f-1+dA_m^2\right)|m\rangle\langle m|
\end{equation}
which turns out to be diagonal. This simplifies the analysis considerably. In fact,
the determinant of $\tilde Q$ is now given by
\begin{equation}
\mathrm{Det(\tilde Q)}=\mathrm{Det}(Q)=\Pi_{k=0}^{d-1}\left(f-1+dA_k^2\right).
\end{equation}
Thereby, the condition $\mathrm{Det}(Q)=0$ implies $f=1-dA_k^2$
for some $k$. The condition $f\le1$ must also holds. This
condition can be satisfied if $A_k^2\le 1/d$. This rules out the
choice $f=1-dA_{d-1}^2$ where $A_{d-1}$ is the the largest
amplitude. The matrix $Q$ (and $\tilde Q$) must also be positive
semidefinite, this can be guaranteed if all the principal minors
of $\tilde Q$ are non-negative, that is
\begin{equation}
\Pi_{k=0}^{n}\left(f-1+dA_k^2\right)\ge0~~~~\forall~n=0,\dots, d-1.
\end{equation}
This condition implies that
\begin{equation}
f\ge 1-dA_k^2~~~~\forall~k=0,\dots, d-1
\end{equation}
and can be satisfied iff
$f=1-d\times\mathrm{min}\{A_k^2\}_{k=0,\dots, d-1}$.  Thus, the
optimal average failure probability $F_{min}$ is given by
\begin{equation}
F_{min}=1-S_{max}=1-d\times A_{min}^2.
\label{Main Result}
\end{equation}
where $A_{min}$ is the smallest coefficient in the state Eq.
(\ref{Initial state}).  The set of states $\{|\phi_l\rangle\}$ can
be readily found. Noting that $Q=F^\dagger\tilde
QF=F^\dagger\tilde A^\dagger \tilde AF=F^\dagger \tilde A^\dagger
FF^\dagger \tilde AF$ we obtain
\begin{equation}
|\phi_l\rangle=\frac{1}{\sqrt{d}}\sum_{k=0}^{d-1}A^l_k|a_k\rangle=
\frac{1}{\sqrt{d}}\sum_{k=0}^{d-1}\Big(\sum_{m=0}^{d-1}\exp(\frac{2\pi i}{d}m(l-k))\sqrt{A_m^2-A_{min}^2}\Big)|a_k\rangle.
\end{equation}

\section{Fidelity of conclusive state teleportation}
\label{sec:three}

With the results of the previous section we can now state
precisely the  protocol for conclusive quantum state
teleportation. Our starting point are the identity Eq. (\ref{ID2})
and the definition of states $|\nu_l\rangle$ of Eq. (\ref{Nu}).
The standard teleportation protocol consists in measuring the
states of particles two and three and communicating the outcomes
$(l,k)$ of these measurement to the receiver. This projects
particle one to the state $Z^{d-l}X^k|\psi\rangle$ from which the
state $|\psi\rangle$ to be teleported can be obtained via applying
the operator $X^{d-k}Z^{l}$. However, the states ${|\nu_l\rangle}$
cannot be distinguished with certainty affecting the overall
performance of the process. At this stage enters optimal
conclusive quantum state discrimination. Before measuring
particles two and three the unitary transformation $U$ Eq.
(\ref{Unitarydiscrimination}) is applied onto particle two. This
leads to the join state
\begin{equation}
U_2GXOR_{23}|\Psi\rangle_{12}\otimes|\psi\rangle_3=\frac{1}{d}
\sum_{l,k=0}^{d-1}Z_1^{d-l}X_1^k|\psi\rangle_1\otimes
(\sqrt{S_{max}}|e_l\rangle_2+|\phi_l\rangle_2)
\otimes|k\rangle_{3}.
\label{SDID4}
\end{equation}
Measurements on particles two and three project the particle one to the state
 \begin{equation}
|\Psi_{l,k}^{\cal U}\rangle=
\frac{1}{d}Z^{d-l}X^{k}|\psi\rangle
\label{Cond1}
\end{equation}
with probability $S_{max}$ and to the state
\begin{equation}
|\Psi_{s,k}^{\cal A}\rangle=
\frac{1}{d\sqrt{d}}
\big(\sum_{l=0}^{d-1}A_s^lZ^{d-l}X^{k}\big)|\psi\rangle
\label{Cond2}
\end{equation}
with probabilities $1-S_{max}$. The state $|\Psi_{l,k}^{\cal
U}\rangle$ Eq.  (\ref{Cond1}) is associated with the conclusive
events in the discrimination of states and clearly leads to the
perfect teleportation of the state $|\psi\rangle$. However, the
state $|\Psi_{s,k}^{\cal A}\rangle$ Eq. (\ref{Cond2}) leads to a
failure of the process. Nevertheless, this state has some fidelity
with respect to the state $|\psi\rangle$ to be teleported. The
average fidelity $F$ of teleportation is given by
\begin{equation}
F=dA_{min}^2+\sum_{s,k=0}^{d-1}\int d\psi|\langle\psi|\Psi_{s,k}\rangle|^2
\end{equation}
where the states $|\Psi\rangle_{s,k}$ are the states of particle
one after the teleportation protocol has been carried out for the
particular pair of outcomes $(s,k)$ and the integral is performed
over the entire Hilbert space . In the case that the
teleportation protocol is interrupted after the measurement of
particles two and three, that is
$|\Psi\rangle_{s,k}=|\Psi\rangle_{s,k}^{\cal A}$ we obtain for the
average fidelity
\begin{equation}
F=dA_{min}^2+\sum_{n,k=0}^{d-1}(A_{n+k}^2-A_{min}^2)\int d\psi |\langle n+k|\phi\rangle|^2|\langle n|\phi\rangle|^2
\label{Fidelity}
\end{equation}
where the states $\{|n\rangle\}$ for $n=0\dots d-1$   for a basis
for the Hilbert space of particle one. The integral entering in
Eq. (\ref{Fidelity}) is equal to $(\delta_{n,k}+1)/d(d+1)$.
Thereby, the average fidelity of teleportation becomes
\begin{equation}
F_0=\frac{1}{d}+(d-1)A_{min}^2.
\end{equation}
The  fidelity can be increased by using the  information available
about the outcomes of the measurements carried out on particles
two and three. The state of Eq. (\ref{Cond2}) can be cast in the
form
\begin{equation}
|\Psi_{s,k}^{\cal A}\rangle=
\frac{1}{\sqrt{d}}
\big(\sum_{l=0}^{d-1}A_s^l\exp(-\frac{2\pi i}{d}lk)Z^{d-l}\big)|\psi\rangle
\label{Cond3}
\end{equation}
by appliying onto particle one the operator  $X^{d-k}$ conditional
on the outcome $k$ of the measurement of particle three. This
state leads to the following fidelity
\begin{equation}
F_1=\frac{2+d(d-1)A_{min}^2}{d+1}.
\end{equation}
which is clearly larger than $F_0$. A further  increase in the
average fidelity can be achieved by observing that the
distribution $(A_s^l)^2$ has its maximum at $s=l$. This suggests
to complete the protocol for unambiguous state teleportation by
applying the operator $Z^s$ onto particle one conditional on the
outcome $s$ of the measurement carried out on particle three. In
this case the average fidelity becomes
\begin{equation}
F_2=\frac{1}{d+1}\big(2+d(d-1)A_{min}^2\big)+\frac{1}{d+1}\sum_{n\ne r}\sqrt{A_{n}^2-A_{min}^2}\sqrt{A_{r}^2-A_{min}^2}.
\label{F2}
\end{equation}
Clearly, $F_0\le F_1\le F_2$ for all $A_{min}~\in~[0,1/\sqrt{d}]$.
In what follows we will show this result to be optimal. This can
be done by noting that the states $|\phi_l\rangle$ can be cast in
the form
\begin{equation}
\tilde F|\phi_l\rangle=\tilde Z^l\sum_{n=0}^{d-1}\sqrt{A_n^2-A_{min}^2}|a_n\rangle
\end{equation}
where the operators $\tilde F$ and $\tilde Z$ act now  on the
subspace ${\cal A}$. Thereby the transformation $U$ is replaced by
$\tilde FU$. This resembles the definition of the states
$|\nu_l\rangle$ and suggests that the states $F|\phi_l\rangle$
originates in a quantum channel $|ch\rangle_{12}$ of the form
\begin{equation}
|ch\rangle_{12}=\sum_{n=0}^{d-1}
\sqrt{\frac{A_n^2-A_{min}^2}{1-dA_{min}^2}}
|a_n\rangle_1\otimes|a_n\rangle_2.
\end{equation}
Thus, conclusive state teleportation can be  described as starting
with a coherent superposition of two quantum channels,
i.\thinspace e.
\begin{equation}
|\psi\rangle_{12}=\sqrt{dA_{min}^2}\sum_{n=0}^{d-1}\frac{1}{\sqrt{d}}|e_n\rangle_1\otimes|e_n\rangle_2+
\sqrt{1-dA_{min}^2}\sum_{n=0}^{d-1}
\sqrt{\frac{A_n^2-A_{min}^2}{1-dA_{min}^2}}
|a_n\rangle_1\otimes|a_n\rangle_2.
\label{Totalchannel}
\end{equation}
The first term at the r.h.s. of Eq. (\ref{Totalchannel})
correspond to a perfectly entangled state in the subspace ${\cal
H}$. This part of the channel is responsible for the events in
which teleportation success with unity fidelity. The second term
at the r.h.s. of Eq. (\ref{Totalchannel}) describes the ambiguous
events which lead to a failure of the teleportation. This term
corresponds to a non-maximally entangled state in the subspace
${\cal A}$ and is formed by the superposition of only $d-1$
states. The protocol for conclusive quantum state teleportation is
easily obtained from
\begin{equation}
GXOR_{2,3}|\psi\rangle_{12}\otimes|\psi\rangle_{3}=\frac{1}{d}
\sum_{l,k=0}^{d-1}Z_1^{d-l}X_1^k|\psi\rangle_1\otimes(
\sqrt{dA_{min}^2}|e_{l}\rangle_{2}+
\sqrt{1-dA_{min}^2}|\nu_{l}\rangle_2)\otimes|k\rangle_3,
\label{dst}
\end{equation}
where
\begin{equation}
|\nu_{l}\rangle=\tilde
Z^l\sum_{n=0}^{d-1}\sqrt{\frac{A_n^2-A_{min}^2}{1-dA_{min}^2}}|a_n\rangle.
\end{equation}

Now we can recall Banaseck's result \cite{Banaszek} concerning the
maximal average fidelity of teleportation
\begin{equation}
F_B\le\frac{1}{d+1}[1-(\sum_{k=0}^{d-1}t_k)^2]
\label{Ban}
\end{equation}
through a quantum channel
$|ch\rangle_{12}=\sum_{k=0}^{d-1}t_k|k\rangle_1\otimes
|k\rangle_2$. Inserting  the coefficients of the quantum channel
Eq. (\ref{Totalchannel}) into the previous definition Eq.
(\ref{Ban}) the fidelity for this channel is given by $F_2$ Eq.
(\ref{F2}). Therefore, the protocol for unambiguous state
teleportation achieves the maximal possible fidelity.

\section{Conclusions}
The problem of teleporting the state of a d-dimensional quantum
system through a quantum channel corresponding to a non-maximally
entangled state is directly related to that of distinguishing
between a set of d non-orthogonal states. In the teleportation
process a generalized control-not gate is used to map the
entangled state of the sender into an unentangled state. The
sender then measures the state which is in general non-orthogonal
and transmits the outcome to the receiver. The receiver then
proceeds to use this information to choose among a set of
transformations that must be applied to his state in order to
recover the teleported state. The states fidelity depends on the
discrimination scheme used by the sender to distinguish its state.
The optimal conclusive discrimination protocol proposed by Sun et.
al. \cite{Sun} is at the center of the teleportation procedure
presented here.

The optimal conclusive discrimination scheme is based on the idea
of mapping the non-orthogonal state onto a set of orthogonal
states in a probabilistic fashion. When the map is successful the
state can then be distinguished with certainty thus leading to
perfect teleportation. The failure probability of the mapping
procedure is responsible for a reduction in the fidelity of the
teleported state. We conclude that can this be visualized in the
following way: the non-maximally entangled quantum channel is a
coherent superposition of two channels, one allowing for perfect
teleportation because the channel corresponds to a maximally
entangled state related to a successful map in the discrimination
procedure and a non-maximally entangled channel living in a
subspace of the Hilbert space. The truncated channel is generated
by the failure probability of the map and teleportation through
this channel leads to a state with reduced fidelity. Linear
independence of the set of non-orthogonal states is a crucial
factor in the scheme. The success of the of the procedure depends
strongly on the number of states which are linearly independent.
If no states are dependent then the only source of error is the
discrimination scheme itself. The dependent states cause further
errors in the scheme because these states cannot be distinguished.
Obviously when all the states are linearly dependent the scheme
fails completely. In the protocol proposed here, the average
fidelity when all the non-orthogonal states are linearly
independent is optimal, i.e. it achieves the maximal average
fidelity possible for a teleportation procedure. Therefore we know
with certainty that any further local operation would only
decrease its performance.

We are currently investigating other possible applications of the
quantum channel (\ref{Totalchannel}) including the teleportation
of unitary evolutions which allow for the remote implementation of
quantum gates. We are also interested in extending the scheme to
continuous variables.
\section{Acknowledgements}
This work was supported under the auspices of the Office of Naval Research Grants Nos.
N00014-00-1-0575 and N00014-00-1-0578. A. D. acknowledges support from Fondecyt under Grant No.
1030671. L. R. thanks Fondecyt for support under Grants No. 1010010, No. 1030671 and UdeC P.I. 96011016-12.
 I. F.-G. would like to thank Consejo Nacional de Ciencia y Tecnologia (Mexico) Grant
no. 115569/135963 for financial support and the Information
Physics group at the University of New Mexico for their
hospitality.

\end{document}